\documentclass[twocolumn,english]{revtex4-1}
\usepackage[T1]{fontenc}
\usepackage[latin9]{inputenc}
\usepackage{amsmath}
\usepackage{amssymb}
\usepackage{graphicx}
\usepackage{babel}

\usepackage{color}

\usepackage{ulem}

\begin{document}
\title{Cavity Quantum Electrodynamics Effects with Nitrogen Vacancy Center Spins in Diamond and Microwave Resonators at Room Temperature}

\author{Yuan Zhang}
\email{yzhuaudipc@zzu.edu.cn}
\address{Henan Key Laboratory of Diamond Optoelectronic Materials and Devices, Key Laboratory of Material Physics, Ministry of Education,School of Physics and Microelectronics, Zhengzhou University, Daxue
Road 75, Zhengzhou 450052 China}

\author{Qilong Wu}
\address{Henan Key Laboratory of Diamond Optoelectronic Materials and Devices, Key Laboratory of Material Physics, Ministry of Education,School of Physics and Microelectronics, Zhengzhou University, Daxue
Road 75, Zhengzhou 450052 China}

\author{Shi-Lei Su}
\address{Henan Key Laboratory of Diamond Optoelectronic Materials and Devices, Key Laboratory of Material Physics, Ministry of Education,School of Physics and Microelectronics, Zhengzhou University, Daxue
Road 75, Zhengzhou 450052 China}

\author{Qing Lou}
\address{Henan Key Laboratory of Diamond Optoelectronic Materials and Devices, Key Laboratory of Material Physics, Ministry of Education,School of Physics and Microelectronics, Zhengzhou University, Daxue
Road 75, Zhengzhou 450052 China}

\author{ChongXin Shan}
\email{cxshan@zzu.edu.cn}
\address{Henan Key Laboratory of Diamond Optoelectronic Materials and Devices, Key Laboratory of Material Physics, Ministry of Education,School of Physics and Microelectronics, Zhengzhou University, Daxue
Road 75, Zhengzhou 450052 China}

\author{Klaus M{\o}mer}
\email{moelmer@phys.au.dk}
\address{Aarhus Institute of Advanced Studies, Aarhus University, H{\o}egh-Guldbergs
Gade 6B, DK-8000 Aarhus C, Denmark\\
Center for Complex Quantum Systems, Department of Physics and Astronomy,
Aarhus University, Ny Munkegade 120, DK-8000 Aarhus C, Denmark}

\begin{abstract}
Cavity quantum electrodynamics (C-QED) effects, such as Rabi splitting, Rabi oscillations and superradiance, have been demonstrated with nitrogen vacancy center spins in diamond in microwave resonators at cryogenic temperature. In this article we explore the possibility to realize strong collective coupling and the resulting C-QED effects with ensembles of spins at room temperature. Thermal excitation of the individual spins by the hot environment leads to population of collective Dicke states with low symmetry and a reduced collective spin-microwave field coupling. However, we show with simulations that the thermal excitation can be compensated by spin-cooling via optical pumping. The resulting population of Dicke states with higher symmetry implies strong coupling with currently available high-quality resonators and enables C-QED effects with potential applications in quantum sensing and quantum information processing. 
\end{abstract}
\maketitle

\begin{figure}[ht!]
\begin{centering}
\includegraphics[scale=0.45]{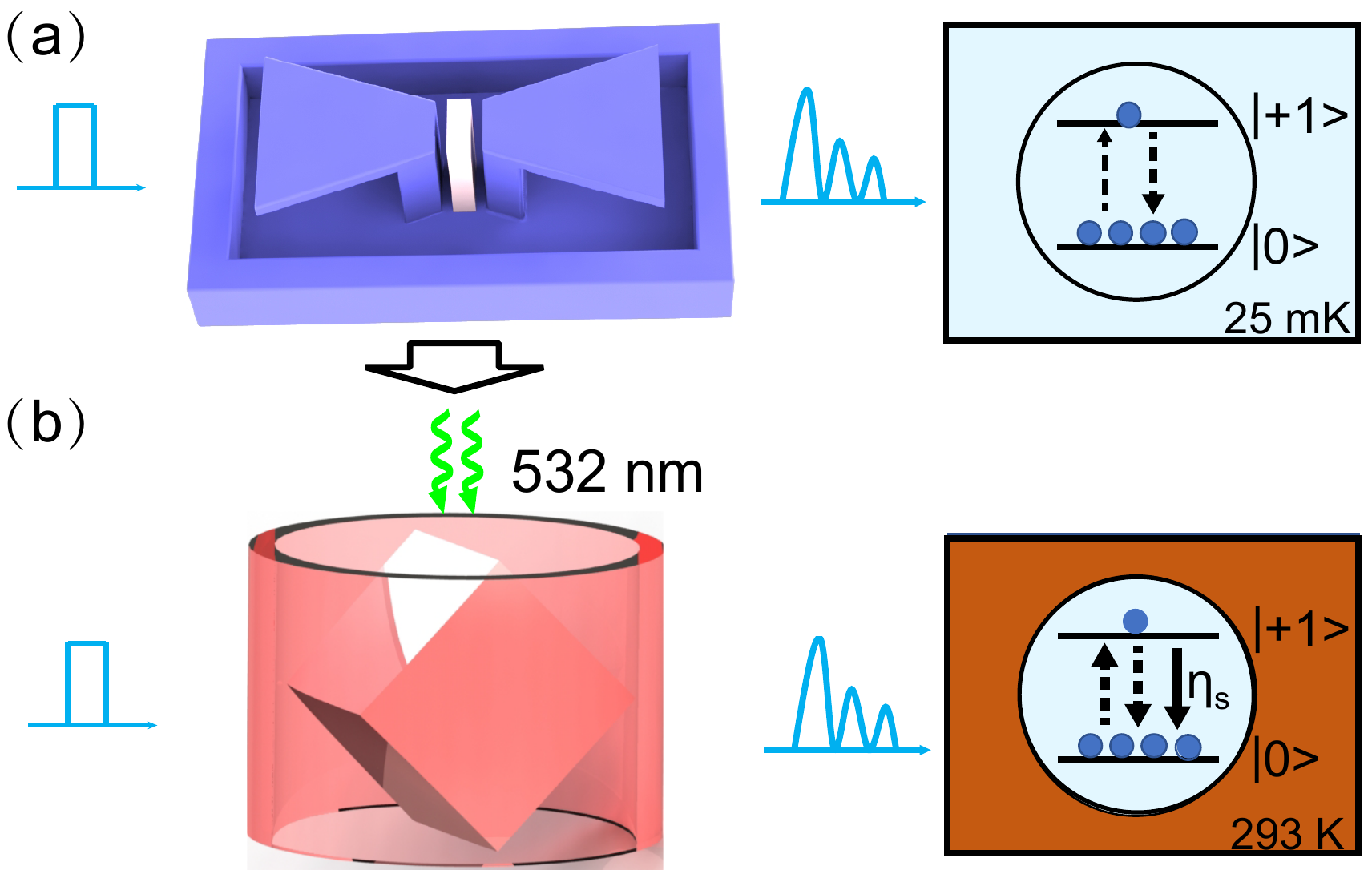}
\par\end{centering}
\caption{\label{fig:system}C-QED systems with a NV center spin ensemble. Panel (a) shows a 3D lumped element resonator operating in a cryogenic environment at
25 mK (left), as used in \citep{AAngerer2018}, where the spin-ensemble is in equilibrium with the cooled environment (right).
Panel (b) shows a sapphire dielectric resonator
operating at room temperature, as used in \citep{JDBreeze2018},
and a diamond  illuminated by 532 nm laser light (left), which counteracts the heating of the spins by the hot environment (right), leading to a higher population in the lower $0$ spin state than in the upper $+1$ spin state. The upwards (downwards) arrows in the panels (a,b) represent the thermal excitation, thermal decay, and the spin-cooling $\eta_s$ via optical pumping. }
\end{figure}

\paragraph{Introduction} Cavity quantum electrodynamics (C-QED) studies the interaction between quantum emitters and cavity photon modes, and addresses both the foundations of quantum mechanics \citep{HMabuchi} and proposals for quantum information processing \citep{JMRaimond} and quantum metrology \citep{JYe}. C-QED effects, such as Rabi splitting \citep{RAmsuss2011,YKubo2010,AAngerer2016}, Rabi-oscillations
\citep{SPutz2014} and superradiance \citep{AAngerer2018},
have been realized  with negatively charged nitrogen vacancy (NV$^{-}$) center spins in diamond in microwave
resonators at cryogenic temperature. Although the NV$^{-}$ center spins have a spin-1 degree of freedom, normally only two of the spin states, say those with projections $m=0$ and $m=+1$, are coupled resonantly to the resonator, and thus the spins can be treated effectively as two-level systems. As illustrated in Fig. \ref{fig:system}(a), at low temperature
the spin ensemble can be highly polarized and can couple strongly with lumped-element microwave resonators with high
Q-factors. 

The low temperature restricts the application of the  C-QED effects in quantum
sensing and quantum information processing, and it would be desirable if the collective coupling could be achieved at room
temperature. In this article, we study
how the temperature affects the quantum state of the spin ensemble,
and we propose to utilize optical pumping to
counteract the thermal heating to achieve strong spin polarization and strong coupling, and to observe the resulting C-QED effects at room temperature with a high-quality dielectric resonator, see Fig. \ref{fig:system}(b). Note that the strong
collective coupling may have been present in the recent room-temperature experiments on continuous wave maser \citep{JDBreeze2018} and dispersive readout of NV$^{-}$ spins \citep{Eisenach2021,Ebel2020} and C-QED effects with the spins in pentacene's lowest triplet states \citep{JDBreeze2017}. 


\paragraph{Nitrogen Vacancy Center Spin Ensemble} Nitrogen vacancy  centers are formed by replacing two adjacent carbon atoms with a nitrogen atom and a vacancy in the diamond lattice. Negatively charged NV$^{-}$ centers have a spin-triplet electronic ground state, which contains three spin states with projection $m=+1,0,-1$ along the axis defined by the nitrogen atom and the vacancy. The $\pm 1$ spin states have higher energy than the $0$ state, and can be split and shifted by a static magnetic field. In this way, the transition between one of the shifted states and the $0$-state can be tuned into resonance with a microwave resonator. Furthermore, we can align the magnetic field along one of four possible nitrogen-vacancy orientations, and ensure that only the associated $0,+1$ spin states couple resonantly with the microwave resonator. These spin states can be viewed as the eigenstates of an effective pseudo 1/2-spin, and $N$ NV$^{-}$ centers with this orientation can be viewed as an ensemble of  $N$ pseudo spins.

\begin{figure}[th]
\begin{centering}
\includegraphics[scale=0.42]{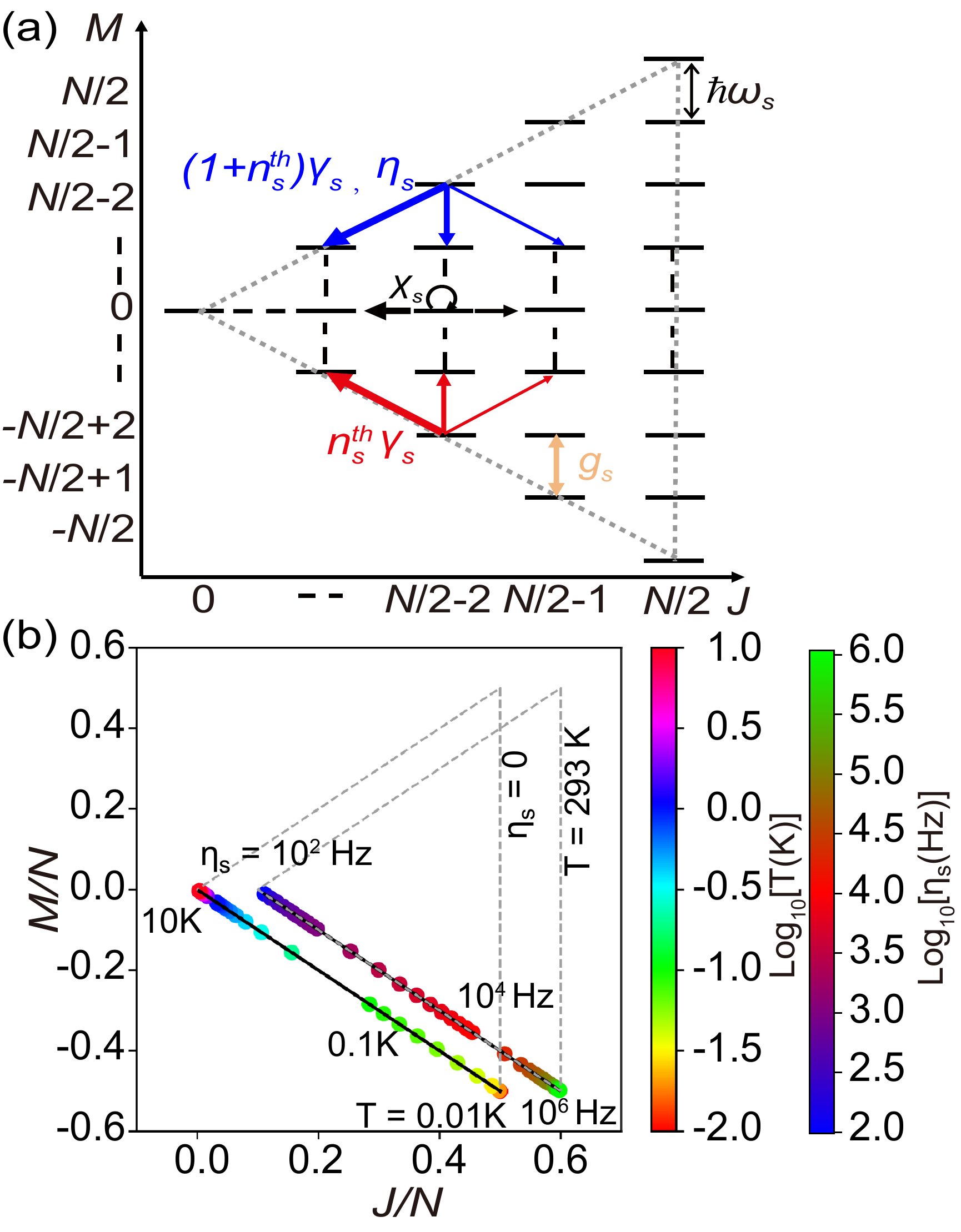}
\par\end{centering}
\caption{\label{fig:dicke} Panel (a) shows the Dicke states of the NV$^{-}$ spin ensemble, and quantum jumps associated with thermal emission and optical spin cooling (blue arrows), thermal excitation  (red arrows), spin dephasing  (black arrows), and collective coupling to the resonator mode (orange double arrows). The symbols in the figure are explained in the text. The transition rates have further dependencies on $J$ and $M$, and the thickness of the arrows represents their relative variation. Panel (b) shows the steady-state Dicke state population for different  temperatures $T$ in the absence of the optical spin cooling
$\eta_{s}=0$ (left), and for different $\eta_{s}$ at room temperature $T=293$ K (right, shifted horizontally for clarity) for parameters compatible with the Rabi-splitting experiment at cryogenic temperature \citep{SPutz2014}. }
\end{figure}

\paragraph{Dicke Representation of Collective States}
The collective coupling of the 1/2-spin ensemble with the quantized electromagnetic field inside the resonator is conventionally described by the so-called Dicke states $\left|J,M\right\rangle $ \citep{RHDicke1054}. The half-integer
or integer number $J\leq J_0=N/2$ refers to the eigenvalues $J(J+1)$ of the collective spin operator $\vec{J}^2$, and the number $M$ within the range $-J\leq M\leq J$ describes the degree of the spin excitation. For each $J$, the Dicke states of different $M$ form a vertical ladder, and the states with different $J$ form the horizontally shifted ladders, constituting a triangular space [Fig. \ref{fig:dicke}(a)]. Due to permutation symmetry, for each pair of $J,M$, there are a total of  $N!(2J+1)/[(J+J_0+1)!(J_0-J)!]$ states, i.e. a degeneracy in the Dicke state energy diagrams, and they couple identically with the cavity field. 

The uniform coupling to the microwave resonator and the individual (but identical) incoherent excitation and decay processes can be consistently and effectively treated as transitions among the Dicke states \citep{YZhang2018,BQBaragiola,NShammah2018}. In Fig. \ref{fig:dicke} (a), we show how the interaction with the thermal environment causes quantum jumps that change
$J$ by $\pm 1$ and $M$ by $0,\pm 1$ (red and blue arrows), while the spin-cooling via optical pumping adds a term $\eta_s$ to the rate $(1+n^{th}_s)\gamma_s$ of downward jumps (blue arrows). Here, $n_s^{th}=[e^{\hbar\omega_s/k_B T}-1]^{-1}$ is the thermal equilibrium phonon excitation at the spin transition energy $\hbar\omega_s$ and temperature $T$ ($k_B$ is Boltzmann's constant). A pure single spin dephasing rate $\chi_s$ is incorporated to describe the phase noise and inhomogeneous broadening of the transition frequencies, and leads to quantum jumps to states with unchanged $M$ and different $J$ (horizontal black arrows), while the
coherent collective coupling with the resonator (with $g_s$ as the coupling to single spin) causes transitions between states of different $M$ and same $J$ (orange arrows). The actual rates also depend on the $J$ and $M$ and for combinatorial reasons the jump probability is larger towards the Dicke states with reduced $J$ (and increased degeneracy) as reflected by the thickness of the lines, see Appendix A in  \citep{YZhang2018} for precise expressions.

In the presence of the thermal processes and the spin cooling, each spin is in a mixed steady state with the upper state population, $p = n_s^{th}\gamma_s/[\eta_s + \gamma_s(1+2n_s^{th})]$. This is equivalent to a distribution on the Dicke states \citep{JWesenberg}, with average values $M,J$ of the quantum numbers defined by the corresponding operator expectation values,   $M=J_0(2p-1)$, and $ J(J+1) = (2p-1)^2J_0(J_0+1) + 6p(p-1)J_0$. For many spins, the relative fluctuations of these quantities are small, and we may thus depict the location of the ensemble steady-state as points in the Dicke state diagram in Fig. \ref{fig:dicke}(b). 

Without the optical spin-cooling $\eta_s=0$, the spin ensemble occupies Dicke states with smaller $J$ for temperature above $T>10$ K, and occupies the states with larger $J$  (lower-right corner) only for $T<0.01$ K, see the dots along the edge of the left triangle in Fig. \ref{fig:dicke}(b).  As the collective spin-resonator coupling scales with $\sqrt{J}$, the spin-ensemble has to be held at very low temperature to ensure the strong collective coupling. In contrast, with sufficient optical spin-cooling (values of $\eta_s$ up to 1 MHz should be achievable in experiments \citep{QWu2021}) to compensate the thermal process, the spin-ensemble can be prepared in Dicke states with larger $J$, see the dots along the rightmost triangle in Fig. \ref{fig:dicke}(b), enabling the strong collective coupling. The spin-ensemble cooling at room temperature as discussed here is fundamentally different from the rapid dephasing-induced and Purcell-enhanced spin cooling via a cavity mode, coupled to a cold environment as proposed in \citep{CJWood}. In the numerical calculations presented below, we shall employ a mean-field approach as in \citep{KDebnath2018,YZhang2021,QWu2021}, and use the Dicke state representation only for visualization and interpretation of the results. 

\paragraph{Quantum Master Equation} To assess the C-QED effects with the NV$^{-}$ spin ensemble-microwave resonator system at room temperature, we must solve the  quantum master equation for the reduced density operator $\hat{\rho}$,
\begin{align}
 & \frac{\partial}{\partial t}\hat{\rho}=-\frac{i}{\hbar}\left[\hat{H}_{c}+\hat{H}_{d}+\hat{H}_{s}+\hat{H}_{s-c},\hat{\rho}\right]\nonumber \\
 & -\kappa_{c}\left[\left(1+n_{c}^{th}\right)\mathcal{D}\left[\hat{a}\right]\hat{\rho}+n_{c}^{th}\mathcal{D}\left[\hat{a}^{+}\right]\hat{\rho}\right]\nonumber \\
 & -\gamma_{s}\left[\left(1+n_{s}^{th}\right)\sum_{j}\mathcal{D}\left[\hat{\sigma}_{j}^{12}\right]\hat{\rho}+n_{s}^{th}\sum_{j}\mathcal{D}\left[\hat{\sigma}_{j}^{21}\right]\hat{\rho}\right]\nonumber \\
 & -\eta_{s}\sum_{j}\mathcal{D}\left[\hat{\sigma}_{j}^{12}\right]\hat{\rho}-2\chi_{s}\sum_{j}\mathcal{D}\left[\hat{\sigma}_{j}^{22}\right]\hat{\rho},\label{eq:meq-1}
\end{align}
where  $\hat{H}_{c}=\hbar\omega_{c}\hat{a}^{\dagger}\hat{a}$
describes the microwave resonator with frequency $\omega_{c}$, photon creation $\hat{a}^{\dagger}$ and annihilation operator $\hat{a}$,
$\hat{H}_{d}=\hbar\Omega\sqrt{\kappa_{1}}\hat{a}e^{i\omega_d t}+h.c.$
describes the driving of the resonator by a microwave probe field
with amplitude $\Omega$, frequency $\omega_{d}$ and the
resonator coupling coefficient $\sqrt{\kappa_{1}}$ ($\kappa_{1}$ is the resonator loss rate due to the same coupling). The spin Hamiltonian $\hat{H}_{s}=\hbar\omega_{s}\sum_{j=1}^{N}\hat{\sigma}_{j}^{22}$
involves the projection operator $\hat{\sigma}_{j}^{22}$ on the upper level  of the $j^{th}$ 
spin, and the spin-resonator interaction Hamiltonian $\hat{H}_{s-c}=\hbar g_{s}\left(\hat{a}^{\dagger}\sum_{j}\hat{\sigma}_{j}^{12}+\sum_{j}\hat{\sigma}_{j}^{21}\hat{a}\right)$ depends on  the lowering $\hat{\sigma}_{j}^{12}$ and raising  operators $\hat{\sigma}_{j}^{21}$ of the spin transitions. 

The second line of Eq.~(\ref{eq:meq-1}) describes the thermal emission and excitation
of the resonator with a rate $\kappa$ and a thermal equilibrium photon number $n_{c}^{th}=\left[e^{\hbar\omega_{c}/k_{B}T}-1\right]^{-1}$
at temperature $T$. Here, the Lindblad superoperator for any operator $\hat{o}$ is defined as $\mathcal{D}\left[\hat{o}\right]\hat{\rho}=\frac{1}{2}\left(\hat{o}^{\dagger}\hat{o}\hat{\rho}+\hat{\rho}\hat{o}^{\dagger}\hat{o}\right)-\hat{o}\hat{\rho}\hat{o}^{\dagger}$. The third and fourth line describe the thermal process, the optical spin-cooling and the dephasing of the spins, as already discussed in Fig. \ref{fig:dicke}. Note that the single dephasing rate modeling of the inhomogeneous broadening is expected to offer a good approximation to the general behavior of the spin ensemble, which can be verified with a more delicate modelling with sub-ensembles of different transition frequencies \citep{AAngerer2018,QWu2021,ABychek}.

To solve Eq. \eqref{eq:meq-1},
we follow the mean-field approach \citep{KDebnath2018,YZhang2021}
(also known as the cluster-expansion method \citep{HAMLeymann2014}) to derive the equation
$\frac{\partial}{\partial t}\left\langle \hat{o}\right\rangle =\mathrm{tr}\left\{ \left(\frac{\partial}{\partial t}\hat{\rho}\right)\hat{o}\right\} $
for the mean value $\left\langle \hat{o}\right\rangle =\mathrm{tr}\left\{ \hat{\rho}\hat{o}\right\} $
of any operator $\hat{o}$, and
truncate the resulting  equation hierarchy by approximating the mean values of products  of many operators with those of fewer operators. To automatically derive the mean-field equations to second order and solve these equations numerically, we make use of the QuantumCumulant.jl package \citep{DPlankensteiner2021}, and present our code and the resulting equations in  App.
\ref{sec:JuliaCode} and \ref{sec:MeanEquations}.

The obtained equations
involve first-order mean quantities, i.e. the resonator
field amplitude $\left\langle \hat{a}\right\rangle $, the spin
coherence $\left\langle \hat{\sigma}_{1}^{12}\right\rangle $ and
the spin upper-level population $\left\langle \hat{\sigma}_{1}^{22}\right\rangle $,
and second-order mean quantities, i.e. the resonator photon
number $\left\langle \hat{a}^{+}\hat{a}\right\rangle $, photon
correlation $\left\langle \hat{a}\hat{a}\right\rangle $, and the spin-photon
correlations $\left\langle \hat{a}^{+}\hat{\sigma}_{1}^{12}\right\rangle ,\left\langle \hat{a}^{+}\hat{\sigma}_{1}^{22}\right\rangle ,\left\langle \hat{a}\hat{\sigma}_{1}^{12}\right\rangle $,
the spin-spin correlations $\left\langle \hat{\sigma}_{1}^{21}\hat{\sigma}_{2}^{12}\right\rangle ,\left\langle \hat{\sigma}_{1}^{22}\hat{\sigma}_{2}^{12}\right\rangle ,\left\langle \hat{\sigma}_{1}^{12}\hat{\sigma}_{2}^{12}\right\rangle ,\left\langle \hat{\sigma}_{1}^{22}\hat{\sigma}_{2}^{22}\right\rangle $, as well as their complex conjugate quantities. To simulate system with a trillion
spins, we have assumed the same parameters $\omega_s,\gamma_s,n_s^{th},\eta_s,\chi_s$  for all the NV$^{-}$ spins, so that we can utilize the aforementioned quantities (with subscripts $1$ or $1,2$) to represent $\left\langle \hat{\sigma}_{j}^{mn}\right\rangle ,\left\langle \hat{a}^{+}\hat{\sigma}_{j}^{mn}\right\rangle ,\left\langle \hat{a}\hat{\sigma}_{j}^{mn}\right\rangle ,\left\langle \hat{\sigma}_{j}^{mn}\hat{\sigma}_{j'}^{m'n'}\right\rangle $ (with $m,n,m',n'=1,2$), 
which have the same values for all the spins $j$ and spin pairs ($j,j'$).  Although the mean-field approach does not deal directly with the Dicke states, the aforementioned mean values do allow us to calculate the
average of the Dicke state quantum numbers \citep{KDebnath2018,YZhang2021,QWu2021} as, $M=N\left(\left\langle \hat{\sigma}_{1}^{22}\right\rangle -\frac{1}{2}\right)$
and $J=\sqrt{\frac{3}{4}N+N\left(N-1\right)\left(\left\langle \hat{\sigma}_{1}^{21}\hat{\sigma}_{1}^{12}\right\rangle +\left\langle \hat{\sigma}_{1}^{22}\hat{\sigma}_{2}^{22}\right\rangle -\left\langle \hat{\sigma}_{1}^{22}\right\rangle +\frac{1}{4}\right)}$.

\begin{figure}[th!]
\begin{centering}
\includegraphics[scale=0.37]{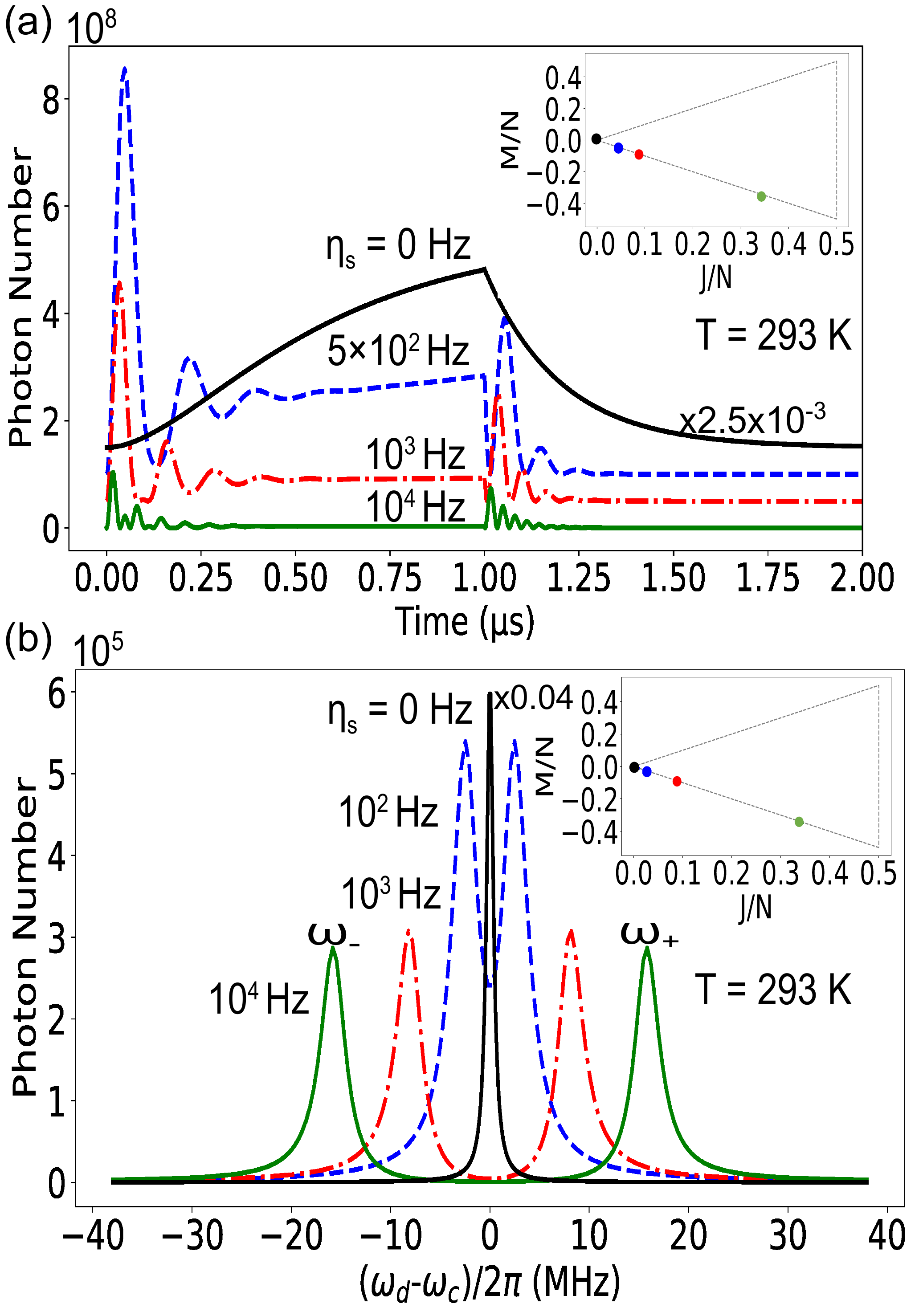}
\par\end{centering}
\caption{\label{fig:Rabi} Cavity field Rabi oscillations and mode splitting  at room temperature. Panel (a) shows the transient evolution of the photon number inside the resonator subject to a short classical pulse for increasing spin-cooling rates $\eta_{s}=0, 5 \times 10^{2},10^{3},10^{4}$ Hz. The results are shifted vertically for clarity. Panel (b) shows the steady-state intra-resonator photon number as function of detuning of the continuous microwave driving field around the resonator frequency for $\eta_{s}=0, 10^{2},10^{3},10^{4}$ Hz, as marked, with peaks from zero detuning and outwards. The inset panels show the Dicke states of the spin-ensemble with increasing $J$ for increasing $\eta_s$. The simulations were done with parameters compatible with the Rabi oscillation experiment at cryogenic temperature \citep{SPutz2014}, see Appendix \ref{sec:para}. Here, we consider a dielectric resonator of same Q-factor at room-temperature (see Ref.\citep{JDBreeze2018,Eisenach2021,Ebel2020,JDBreeze2017}).}
\end{figure}

\paragraph{Rabi Oscillations and Splitting at Room Temperature}

To observe collective Rabi oscillations, we drive the optically cooled spin ensemble with a resonant
square microwave pulse of 1 $\mu$s duration, as in the experiment \citep{SPutz2014},
and  calculate the photon number inside the resonator, see Fig. \ref{fig:Rabi}(a).
We observe that the photon number increases and decays with
an oscillatory amplitude behavior when the driving field switches on and off. Without the optical spin cooling $\eta_s =0$ we observe no Rabi oscillations (black curve), while already for a weak rate $\eta_{s}=5 \times 10^{2}$ (blue dashed line), oscillations appear and  become faster for  higher rates $\eta_{s}= 10^{3},10^{4}$ Hz (red
dash-dotted line and green solid line), which confirms the increased coupling to the more symmetric states with larger  $J$ [inset of Fig. \ref{fig:Rabi}(a)]. We also note that the oscillations end earlier for larger values of $\eta_{s}$.

To observe the coherent splitting of the cavity resonance, we analyze the cavity transmission, proportional to the steady-state photon number inside the resonator, as function of the driving field frequency [Fig. \ref{fig:Rabi}(b)]. For no optical spin cooling $\eta_s=0$, the cavity shows a conventional single peak, while a dip, known as the Fano-effect, appears already  for a weak rate $\eta_{s}=10^2$ (blue dashed line). The dip evolves into two  peaks for larger rates $\eta_{s}=10^3,10^4$Hz (red dash-dotted line and  green solid line) due to the higher $J$ values and hence stronger collective spin-resonator coupling  [inset of Fig. \ref{fig:Rabi}(b)].

We show in the insets of Fig. \ref{fig:Rabi}(a,b), that the spin-ensembles occupy states near the lower boundaries of the Dicke state space. Thus, we may approximate these states as occupation number states of quantized harmonic oscillators 
(Holstein-Primakoff approximation \citep{JAGyamfi2019,THolstein,QWu2021}), and treat the spin-resonator system as two  quantized harmonic oscillators coupled with strength $\sqrt{2J}g_s$. For each $J$, diagonalizing the corresponding Hamiltonians yields two hybrid modes, see Appendix \ref{sec:HPApp}. For the resonant case  $\omega_s=\omega_s$, the hybrid modes have the frequencies $\omega_{\pm} = \omega_c\pm \sqrt{2J}g_s$, and lead to the two peaks split by $2\sqrt{2J}g_s$ as in Fig. \ref{fig:Rabi}(b). Since the value of $J$ does not change much due to the driving of the resonator, we can determine the Rabi peak positions by our analytical expression for the mean value of $J$, as function of the thermal and optical spin-cooling rates. In Appendix \ref{sec:sensing}, we show that by measuring the positions and heights of the Rabi peaks, as in Fig. \ref{fig:Rabi}(b), we can sense the spin transition frequency and hereby related quantities, such as a magnetic field.

\begin{figure}[th!]
\begin{centering}
\includegraphics[scale=0.50]{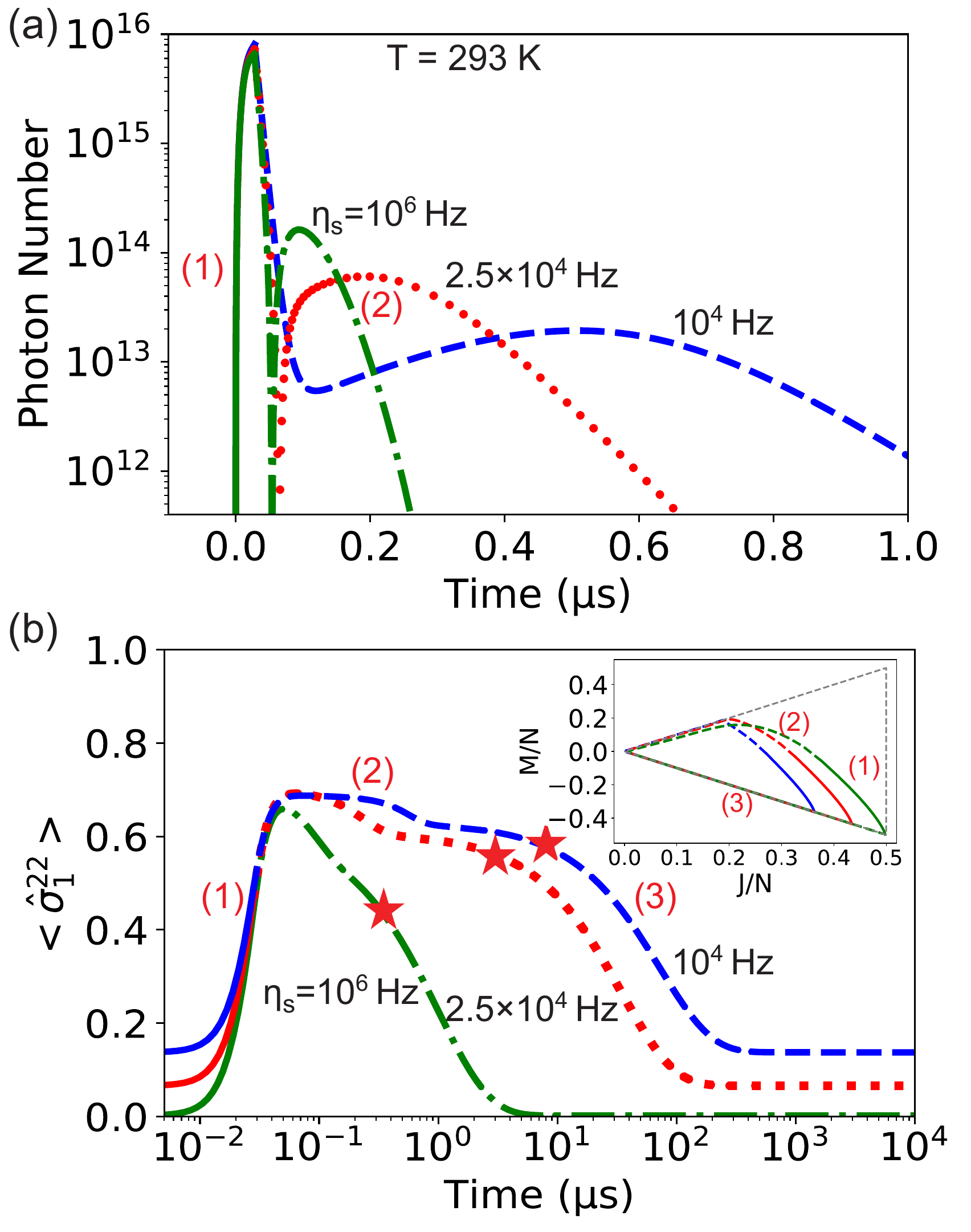}
\par\end{centering}
\caption{\label{fig:superradiance} Stimulated superradiance pulse of NV$^{-}$ spins at room temperature. Panels (a,b) show the evolution of the mean intra-resonator photon number (a), and the population of the upper spin state (b) (the inset shows the  corresponding evolution of the spin-ensemble Dicke state mean quantum numbers) during driving (1), stimulated superradiant emission (2) and spin re-cooling (3), for the optical spin-cooling rate $\eta_{s}=10^4,2.5\times10^4,10^{6}$ Hz, as marked. The red stars in panel (b) mark the end of the superradiant phase. In the simulations, we assume parameters compatible with the stimulated superradiance pulse experiment at cryogenic temperature
\citep{AAngerer2018}, see Appendix \ref{sec:para}. Here, we consider a dielectric resonator of same Q-factor at room-temperature (see Ref.\citep{JDBreeze2018,Eisenach2021,Ebel2020,JDBreeze2017}).}
\end{figure}

\paragraph{Stimulated Superradiance Pulses at Room Temperature} 
We now study the transient excitation and subsequent decay of the spin-ensemble by the collective coupling to the cavity mode at room temperature [Fig. \ref{fig:superradiance}]. In our simulations, we drive the resonator with a strong microwave field to excite the spin-ensemble [marked by "(1)"], which is initially cooled by optical pumping to Dicke states with different degrees of symmetry. Then, we switch off the microwave driving, and the spin ensemble dephases and decays towards lower excitation states with reduced values of $J$, leading to the radiation pulse  [marked by "(2)"]. Finally, the optical spin-cooling and dephasing  re-initialize the spin-ensemble into states with larger $J$ [marked by "(3)"]. The simulated photon number dynamics [Fig. \ref{fig:superradiance}(a)] is similar to the  results observed in experiments at cryogenic temperature \citep{AAngerer2018}, but it occurs here with spin cooling in an apparatus maintained at room temperature. We note that due to the strong spin dephasing in our system, right after the emission [Fig. \ref{fig:superradiance}(b)] the population of the upper spin state decays first to some finite value around 0.5, equivalent to  Dicke states with low symmetry and small $J$. Our results also show that the radiation pulses become stronger and shorter for larger values of $\eta_s$. Because of the involvement of the stimulated emission, the large number of photons and the exploration of collective Dicke states, the radiation is qualified as stimulated superradiance. With the optical pumping, after the radiation pulse the spin ensemble is cooled  to the desired Dicke states with higher symmetry within $4 {\rm \mu s}$ to $300 {\rm \mu s}$ for $\eta_s=10^6$ Hz to $10^4$ Hz, which is much faster than the thermal cooling within a cryogenic environment. 

\paragraph{Conclusions} 
In summary, we have demonstrated theoretically that optical pumping can counteract thermalization of spin states in a room temperature environment and prepare and maintain an NV$^{-}$ spin-ensemble in Dicke states with high symmetry and strong collective coupling to a microwave resonator. Using parameters compatible with existing experimental setups, we show that C-QED effects, such as collective Rabi oscillations, Rabi splitting and stimulated superradiance can be realized at room temperature. The realization of strong spin polarization at room temperature and the resulting collective coupling may enable further applications and studies of quantum dynamics such as quantum memories \citep{IDiniz2011} and self-stimulated
spin echos \citep{KDebnath2020} in ambient  environments. 

\begin{acknowledgments}
We acknowledge Hao Wu for helpful discussions on the physics and the manuscript. This work was supported by the National Natural Science Foundation of China through the project No. 12004344 and the project No. 62027816, as well as the Danish National Research Foundation through the Center of Excellence for Complex Quantum Systems (Grant agreement No. DNRF156).
\end{acknowledgments}

\section*{Author contributions}
Yuan Zhang and Qilong Wu contribute equally to this work. All the authors contribute to the writing of the manuscript.

\appendix

\section{Julia Code to Derive and Solve Mean-Field Equations\label{sec:JuliaCode}}

\begin{figure}
\begin{centering}
\includegraphics[scale=0.45]{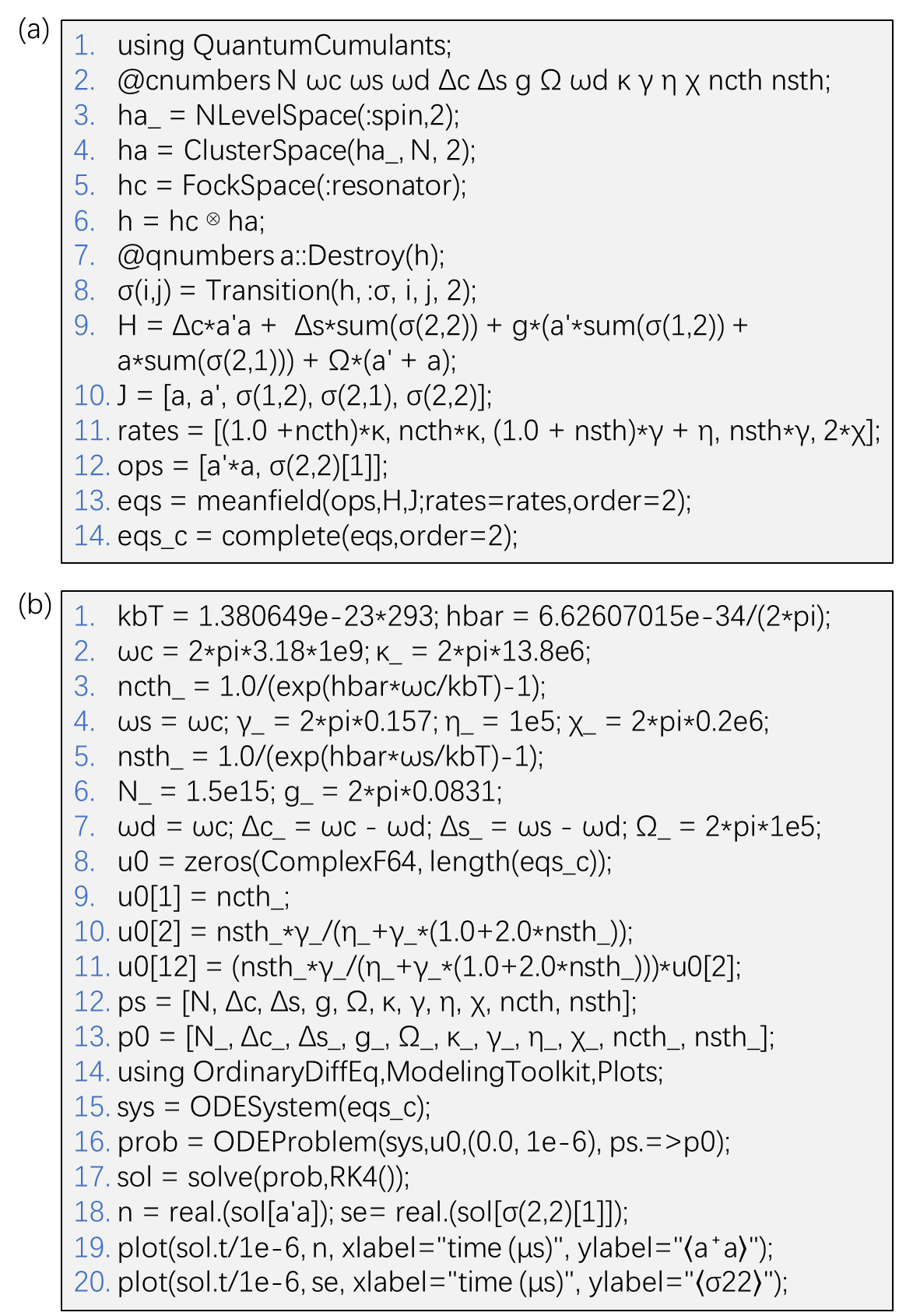}
\par\end{centering}
\caption{\label{fig:code}Julia codes used to derive and solve our mean field equations. Panel (a) shows
the code to derive the symbolic equations for the second order mean-field quantities
from the master equation (\ref{eq:meq-1}). We work in a frame rotating with the driving field frequency $\Delta_c =\omega_c -\omega_d$, $\Delta_s = \omega_s -\omega_d$. Panel (b) shows the code to convert the symbolic equations
to the numerical ones, and to solve these equations with numerical
methods. }
\end{figure}

In Fig. \ref{fig:code} (a), we show the Julia code used to derive the second order
mean-field equations from the master equation (\ref{eq:meq-1}). In lines 1 and 2, we import the QuantumCumulants.jl package,
and then define the symbols to represent the complex numbers. In
lines 3 and 4, we define the Hilbert space for the two-level spin, and the special Hilbert space for $N$ two-level spins, which utilizes the symmetry of identical spins and is restricted to the second order. In lines  5 and 6, we define the Fock space for microwave photons in the resonator (a quantized harmonic oscillator),
and the product Hilbert space for the spins-resonator system.
In lines 7 and 8, we define the photon annihilation
operator $a$ ( and its conjugate $a'$), and the projection, raising
and lowering operators of the spins. For example, $\sigma(2,2)[1],\sigma(2,2)[2]$
are the upper-level projection operators of the first and second spin,
and $\sigma(1,2)[1],\sigma(1,2)[2]$ are the lowing operators of the
two representative spin. Note that the lower-level projection operators
$\sigma(1,1)[1],\sigma(1,1)[2]$ are equal to $1-\sigma(2,2)[1],1-\sigma(2,2)[2]$,and are thus not directly
considered in the program. In the line 9, we define
the system Hamiltonian in a frame rotating with the probe field frequency, where the function $\mathrm{sum(\sigma(i,j))=\sigma(i,j)[1]+\sigma(i,j)[2]}$
gives the sum of the spin operators. In lines 10 and 11, we define the operators and the rates used to specify the Lindblad
terms in the master equation. In lines 12 and 13, we specify the photon number operator $a'a$ and the high-level projection
operator of the first spin, and then derive the mean-field equations
in second order for the expectation value of these operators. The program replaces the sum of spins (spin pairs)
by $N$ ($N(N-1)$) times the values for a single spin  (spin pair), and apply the third-order
cumulant approximation $\left\langle \hat{o}\hat{p}\hat{q}\right\rangle \approx\left\langle \hat{o}\right\rangle \left\langle \hat{p}\hat{q}\right\rangle +\left\langle \hat{o}\hat{q}\right\rangle \left\langle \hat{p}\right\rangle +\left\langle \hat{q}\right\rangle \left\langle \hat{o}\hat{p}\right\rangle -2\left\langle \hat{o}\right\rangle \left\langle \hat{p}\right\rangle \left\langle \hat{q}\right\rangle $
(for any operators $\hat{o},\hat{p},\hat{q}$) to account correctly for
expectation values up to second order. In the last line 14, we identify 
other first and second order mean-field quantities appearing
in the derived equations, and we derive the equations for these quantities
to finally form a closed set of equations, see the Appendix \ref{sec:MeanEquations}.

In Fig. \ref{fig:code} (b), we show the Julia code to solve the mean-field
equations. In line 1, we specify the thermal energy at room temperature,
and the Planck constant. In lines 2 and 3, we specify the resonator
frequency, the photon loss rate, and the thermal photon number. In lines
4 and 5, we specify the spin transition frequency, the spin-relaxation
rate, the spin cooling rate, the dephasing rate, and the thermal phonon coupling
rate. In line 6, we specify the number of spins and the spin-resonator
coupling. In line 7, we specify the frequency and amplitude of
driving field, and the frequency detunings. In lines 8-11,
we initialize the mean-field quantities that all vanish except 
the mean photon number $u0[1]$, the equlibrium population of
higher spin level $u0[2]$, and the spin-spin correlation $\left\langle \hat{\sigma}_{1}^{22}\hat{\sigma}_{2}^{22}\right\rangle $ $u0[12]$. In lines 12 and 13, we define
the array of symbolic parameters, and their numerical values. In line
14, we import the OrdinaryDiffEq.jl, ModelingToolkit.jl and
Plots.jl package for ordinary differential equations, the symbolic-numeric
computation and the plot function. In lines 15-17, we convert
the symbolic equations to a form recognized by the ModelingToolkit.jl
module, and we define the ordinary differential equation problem
by giving the equations, the initial values, the simulation time and
the parameters, and finally solve the problem with Rouge-Kutta method.
In lines 18-20, we extract the mean photon number and the
population of the upper spin level, and plot their numerical solutions.
The code given here can be modified to produce all the results shown
in the main text and the appendix. 

\section{Mean Field Equations in Second-order\label{sec:MeanEquations}}

Our analysis reveals that the three first-order mean-field quantities for the resonator
field amplitude $\langle\hat{a}^{\dagger}\rangle$, the spin coherence
$\langle\hat{\sigma}_{1}^{12}\rangle$ and the population of the upper
spin level $\langle\hat{\sigma}_{1}^{22}\rangle$, obey the
following equations:
\begin{align}
 & \partial_{t}\langle\hat{a}^{\dagger}\rangle=i\tilde{\Delta}_{c}^{*}\langle\hat{a}^{\dagger}\rangle+i\Omega\sqrt{\kappa_{1}}+iNg_{s}\langle\sigma_{1}^{21}\rangle,
\end{align}
\begin{align}
 & \partial_{t}\langle\hat{\sigma}_{1}^{12}\rangle=-i\tilde{\Delta}_{s}\langle\hat{\sigma}_{1}^{12}\rangle-ig_{s}\langle\hat{a}\rangle+2ig_{s}\langle\hat{a}\sigma_{1}^{22}\rangle,
\end{align}
\begin{align}
 & \partial_{t}\langle\hat{\sigma}_{1}^{22}\rangle=n_{s}^{th}\gamma_{s}-\left[\eta_{s}+\gamma_{s}(1+2n_{s}^{th})\right]\langle\hat{\sigma}_{1}^{22}\rangle\\
 & +ig_{s}\left(\langle\hat{a}^{\dagger}\hat{\sigma}_{1}^{12}\rangle-\langle\hat{a}\hat{\sigma}_{1}^{21}\rangle\right).
\end{align}
Here and in the following, we work in a frame rotating with the driving field frequency $\omega_d$, and define the frequency detunings  $\Delta_c =\omega_c -\omega_d$, $\Delta_s = \omega_s -\omega_d$, and the complex frequency detunings $\tilde{\Delta}_{c}=\Delta_{c}-i\kappa_{c}/2$
and $\tilde{\Delta}_{s}=\Delta_{s}-i\frac{1}{2}\bigl[\eta_{s}+(1+2n_{s}^{th})\gamma_{s}+2\chi_{s}]$.
We encounter more second order mean-field quantities, i.e. the mean
photon number $\langle a^{\dagger}a\rangle$, the photon correlation
$\langle aa\rangle$, the spin-photon correlations $\langle\hat{a}^{\dagger}\hat{\sigma}_{1}^{12}\rangle,\langle\hat{a}^{\dagger}\hat{\sigma}_{1}^{22}\rangle,\langle\hat{a}\hat{\sigma}_{1}^{12}\rangle$,
and the spin-spin correlations $\langle\hat{\sigma}_{1}^{21}\hat{\sigma}_{2}^{12}\rangle,\langle\hat{\sigma}_{1}^{22}\hat{\sigma}_{2}^{21}\rangle,\langle\hat{\sigma}_{1}^{12}\hat{\sigma}_{2}^{12}\rangle,\langle\hat{\sigma}_{1}^{22}\hat{\sigma}_{2}^{22}\rangle$.
The equations for the photon number and correlation read 

\begin{align}
 & \partial_{t}\langle\hat{a}^{\dagger}\hat{a}\rangle=\kappa_{c}\left(n_{c}^{th}-\langle\hat{a}^{\dagger}\hat{a}\rangle\right)+i\Omega\sqrt{\kappa_{1}}\left(\langle\hat{a}\rangle-\langle\hat{a}^{\dagger}\rangle\right)\nonumber \\
 & +iNg_{s}(\langle\hat{a}\hat{\sigma}_{1}^{21}\rangle-\langle\hat{a}^{\dagger}\hat{\sigma}_{1}^{12}\rangle),
\end{align}
\begin{align}
 & \partial_{t}\langle\hat{a}\hat{a}\rangle=-2i\tilde{\Delta}_{c}\langle\hat{a}\hat{a}\rangle-2i\Omega\sqrt{\kappa_{1}}\langle\hat{a}\rangle-2iNg_{s}\langle\hat{a}\hat{\sigma}_{1}^{12}\rangle.
\end{align}
The equations for the spin-photon correlations read

\begin{align}
 & \partial_{t}\langle\hat{a}^{\dagger}\hat{\sigma}_{1}^{12}\rangle=i\left(\tilde{\Delta}_{c}^{*}-\tilde{\Delta}_{s}\right)\langle\hat{a}^{\dagger}\hat{\sigma}_{1}^{12}\rangle+i\Omega\sqrt{\kappa_{1}}\langle\hat{\sigma}_{1}^{12}\rangle\nonumber \\
 & +ig_{s}\left[\langle\hat{\sigma}_{1}^{22}\rangle+\langle\hat{\sigma}_{1}^{21}\hat{\sigma}_{2}^{12}\rangle(N-1)\right]-ig_{s}\langle\hat{a}^{\dagger}\hat{a}\rangle\nonumber \\
 & +2ig_{s}(\langle\hat{a}^{\dagger}\rangle\langle\hat{a}\hat{\sigma}_{1}^{22}\rangle+\langle\hat{a}\rangle\langle\hat{a}^{\dagger}\hat{\sigma}_{1}^{22}\rangle\nonumber \\
 & +\langle\hat{\sigma}_{1}^{22}\rangle\langle\hat{a}^{\dagger}\hat{a}\rangle-2\langle\hat{a}^{\dagger}\rangle\langle\hat{a}\rangle\langle\hat{\sigma}_{1}^{22}\rangle),
\end{align}

\begin{align}
 & \partial_{t}\langle\hat{a}^{\dagger}\hat{\sigma}_{1}^{22}\rangle=\bigl\{ i\Delta_{c}^{*}-\bigl[\eta_{s}+\gamma_{s}(1+2n_{s}^{th})\bigr]\bigr\}\langle\hat{a}^{\dagger}\hat{\sigma}_{1}^{22}\rangle\nonumber \\
 & +n_{s}^{th}\gamma_{s}\langle\hat{a}^{\dagger}\rangle+i\Omega\sqrt{\kappa_{1}}\langle\hat{\sigma}_{1}^{22}\rangle+ig_{s}(N-1)\langle\hat{\sigma}_{1}^{22}\hat{\sigma}_{2}^{21}\rangle\nonumber \\
 & +ig_{s}\bigl[2\langle\hat{a}^{\dagger}\rangle\langle\hat{a}^{\dagger}\hat{\sigma}_{1}^{12}\rangle+\langle\hat{\sigma}_{1}^{12}\rangle\langle\hat{a}^{\dagger}\hat{a}^{\dagger}\rangle-2\langle\hat{\sigma}_{1}^{12}\rangle\langle\hat{a}a^{\dagger}\rangle^{2}\bigr]\nonumber \\
 & -ig_{s}\bigl[\langle\hat{a}^{\dagger}\rangle\langle\hat{a}\hat{\sigma}_{1}^{21}\rangle+\langle\hat{a}\rangle\langle\hat{a}^{\dagger}\hat{\sigma}_{1}^{21}\rangle\nonumber \\
 & +\langle\hat{\sigma}_{1}^{21}\rangle\langle\hat{a}^{\dagger}\hat{a}\rangle-2\langle\hat{a}^{\dagger}\rangle\langle\hat{a}\rangle\langle\hat{\sigma}_{1}^{21}\rangle\bigr],
\end{align}
\begin{align}
 & \partial_{t}\langle\hat{a}\hat{\sigma}_{1}^{12}\rangle=-i\left(\tilde{\Delta}_{c}^{*}+\tilde{\Delta}_{s}\right)\langle\hat{a}\hat{\sigma}_{1}^{12}\rangle-i\Omega\sqrt{\kappa_{1}}\langle\hat{\sigma}_{1}^{12}\rangle\nonumber \\
 & -ig_{s}\left[\langle\hat{a}\hat{a}\rangle+(N-1)\langle\hat{\sigma}_{1}^{12}\hat{\sigma}_{2}^{12}\rangle\right]\nonumber \\
 & +2ig_{s}(\langle\hat{\sigma}_{1}^{22}\rangle\langle\hat{a}\hat{a}\rangle+2\langle\hat{a}\rangle\langle\hat{a}\hat{\sigma}_{1}^{22}\rangle-2\langle\hat{\sigma}_{1}^{22}\rangle\langle\hat{a}\rangle^{2}).
\end{align}
The equations for the spin-spin correlations read

\begin{align}
 & \partial_{t}\langle\hat{\sigma}_{1}^{21}\hat{\sigma}_{2}^{12}\rangle=-\left[\eta_{s}+\gamma_{s}(1+2n_{s}^{th})+2\chi_{s}\right]\langle\hat{\sigma}_{1}^{21}\hat{\sigma}_{2}^{12}\rangle\nonumber \\
 & +ig_{s}\left(\langle\hat{a}^{\dagger}\hat{\sigma}_{1}^{12}\rangle-\langle\hat{a}\hat{\sigma}_{1}^{21}\rangle\right)\nonumber \\
 & -2ig_{s}(\langle\hat{a}^{\dagger}\rangle\langle\hat{\sigma}_{1}^{22}\hat{\sigma}_{2}^{12}\rangle+\langle\hat{\sigma}_{1}^{12}\rangle\langle\hat{a}^{\dagger}\hat{\sigma}_{1}^{22}\rangle\nonumber \\
 & +\langle\hat{\sigma}_{1}^{22}\rangle\langle\hat{a}^{\dagger}\hat{\sigma}_{1}^{12}\rangle-2\langle\hat{a}^{\dagger}\rangle\langle\hat{\sigma}_{1}^{12}\rangle\langle\hat{\sigma}_{1}^{22}\rangle)\nonumber \\
 & +2ig_{s}(\langle\hat{a}\rangle\langle\hat{\sigma}_{1}^{22}\hat{\sigma}_{2}^{21}\rangle+\langle\hat{\sigma}_{1}^{21}\rangle\langle\hat{a}\hat{\sigma}_{1}^{22}\rangle\nonumber \\
 & +\langle\hat{\sigma}_{1}^{22}\rangle\langle\hat{a}\hat{\sigma}_{1}^{21}\rangle-2\langle\hat{a}\rangle\langle\hat{\sigma}_{1}^{21}\rangle\langle\hat{\sigma}_{1}^{22}\rangle),
\end{align}

\begin{align}
 & \partial_{t}\langle\hat{\sigma}_{1}^{22}\hat{\sigma}_{2}^{21}\rangle=\bigl\{ i\tilde{\Delta}_{s}^{*}-\bigl[\eta_{s}+\gamma_{s}(1+2n_{s}^{th})\bigr]\bigr\}\langle\hat{\sigma}_{1}^{22}\hat{\sigma}_{2}^{21}\rangle\nonumber \\
 & +n_{s}^{th}\gamma_{s}\langle\hat{\sigma}_{1}^{21}\rangle+ig_{s}(\langle\hat{a}^{\dagger}\rangle\langle\hat{\sigma}_{1}^{21}\hat{\sigma}_{2}^{12}\rangle+\langle\hat{\sigma}_{1}^{21}\rangle\langle\hat{a}^{\dagger}\hat{\sigma}_{1}^{12}\rangle\nonumber \\
 & +\langle\hat{\sigma}_{1}^{12}\rangle\langle\hat{a}^{\dagger}\hat{\sigma}_{1}^{21}\rangle-2\langle\hat{a}^{\dagger}\rangle\langle\hat{\sigma}_{1}^{21}\rangle\langle\hat{\sigma}_{1}^{12}\rangle)\nonumber \\
 & -2ig_{s}(\langle\hat{a}^{\dagger}\rangle\langle\hat{\sigma}_{1}^{22}\hat{\sigma}_{2}^{22}\rangle+2\langle\hat{\sigma}_{1}^{22}\rangle\langle\hat{a}^{\dagger}\hat{\sigma}_{1}^{22}\rangle-2\langle\hat{a}^{\dagger}\rangle\langle\hat{\sigma}_{1}^{22}\rangle^{2})\nonumber \\
 & -ig_{s}(\langle\hat{a}\rangle\langle\hat{\sigma}_{1}^{21}\hat{\sigma}_{2}^{21}\rangle+2\langle\hat{\sigma}_{1}^{21}\rangle\langle\hat{a}\hat{\sigma}_{1}^{21}\rangle\nonumber \\
 & -2\langle\hat{a}\rangle\langle\hat{\sigma}_{1}^{21}\rangle^{2})+ig_{s}\langle\hat{a}^{\dagger}\hat{\sigma}_{1}^{22}\rangle,
\end{align}

\begin{align}
 & \partial_{t}\langle\hat{\sigma}_{1}^{12}\hat{\sigma}_{2}^{12}\rangle=-i2\tilde{\Delta}_{s}\langle\hat{\sigma}_{1}^{12}\hat{\sigma}_{2}^{12}\rangle-i2g_{s}\langle\hat{a}\hat{\sigma}_{1}^{12}\rangle\nonumber \\
 & +i4g_{s}(\langle\hat{a}\rangle\langle\hat{\sigma}_{1}^{22}\hat{\sigma}_{2}^{12}\rangle+\langle\hat{\sigma}_{1}^{12}\rangle\langle\hat{a}\hat{\sigma}_{1}^{22}\rangle\nonumber \\
 & +\langle\hat{\sigma}_{1}^{22}\rangle\langle\hat{a}\hat{\sigma}_{1}^{12}\rangle-2\langle\hat{a}\rangle\langle\hat{\sigma}_{1}^{12}\rangle\langle\hat{\sigma}_{1}^{22}\rangle),
\end{align}

\begin{align}
 & \partial_{t}\langle\hat{\sigma}_{1}^{22}\hat{\sigma}_{2}^{22}\rangle=-2\bigl[\eta_{s}+\gamma_{s}(1+2n_{s}^{th})\bigr]\langle\hat{\sigma}_{1}^{22}\hat{\sigma}_{2}^{22}\rangle\nonumber \\
 & +2n_{s}^{th}\gamma_{s}\langle\hat{\sigma}_{1}^{22}\rangle+i2g_{s}(\langle\hat{a}^{\dagger}\rangle\langle\hat{\sigma}_{1}^{22}\hat{\sigma}_{2}^{12}\rangle+\langle\hat{\sigma}_{1}^{12}\rangle\langle\hat{a}^{\dagger}\hat{\sigma}_{1}^{22}\rangle\nonumber \\
 & +\langle\hat{\sigma}_{1}^{22}\rangle\langle\hat{a}^{\dagger}\hat{\sigma}_{1}^{12}\rangle-2\langle\hat{a}^{\dagger}\rangle\langle\hat{\sigma}_{1}^{12}\rangle\langle\hat{\sigma}_{1}^{22}\rangle-\langle\hat{a}\rangle\langle\hat{\sigma}_{1}^{22}\hat{\sigma}_{2}^{21}\rangle\nonumber \\
 & -\langle\hat{\sigma}_{1}^{21}\rangle\langle\hat{a}\hat{\sigma}_{1}^{22}\rangle-\langle\hat{\sigma}_{1}^{22}\rangle\langle\hat{a}\hat{\sigma}_{1}^{21}\rangle+2\langle\hat{a}\rangle\langle\hat{\sigma}_{1}^{21}\rangle\langle\hat{\sigma}_{1}^{22}\rangle).
\end{align}
By examining the mean-field equations, we find that they
depend also on the quantities $\left\langle a\right\rangle ,\langle\hat{\sigma}_{1}^{21}\rangle,\langle\hat{a}^{+}\hat{a}^{+}\rangle,\langle\hat{a}\hat{\sigma}_{1}^{21}\rangle,\langle\hat{a}\hat{\sigma}_{1}^{22}\rangle,\langle\hat{a}^{\dagger}\hat{\sigma}_{1}^{21}\rangle,\langle\hat{\sigma}_{1}^{22}\hat{\sigma}_{2}^{12}\rangle,\langle\hat{\sigma}_{1}^{21}\hat{\sigma}_{2}^{21}\rangle$.
Since these quantities are the complex conjugate of quantities appearing
already, i.e. $\left\langle a\right\rangle =\left\langle a^{\dagger}\right\rangle ^{*},\langle\hat{\sigma}_{1}^{21}\rangle=\langle\hat{\sigma}_{1}^{12}\rangle^{*}$,
we do not need to consider separate equations for these terms.

\section{ Simulations Parameters \label{sec:para}}

In this appendix, we summarize the parameters used for the simulations in Fig. 3 and Fig. 4 in the main text. For Fig.3, we utilize the parameters compatible with the Rabi oscillation experiment at cryogenic temperature \citep{SPutz2014}: the  resonator has the frequency $\omega_{c}=2\pi\times2.69$ GHz, the photon loss rate  $\kappa_{c}=2\pi\times0.8$ MHz ($\kappa_1=\kappa_{c}/2$), and the coupling with single spin $g_{s}=2\pi\times12$ Hz; the spins have the transition frequency $\omega_{s}=\omega_{c}$, the spin-lattice relaxation rate $\gamma_{s}=2\pi\times0.157$ Hz, and the dephasing $\chi_{s}=2\pi\times {2.6}$ MHz. In addition, the number of spins is $N=2.5\times10^{12}$. For Fig. 3(a), the driving microwave field drives resonantly the resonator ($\omega_d = \omega_c$) with the amplitude $\Omega = 2 \pi \times 10^{8}$ Hz$^{-1/2}$ Hz$^{-1/2}$. For Fig. 3(b), the driving microwave field has the amplitude $\Omega = 2 \pi \times 10^{6}$ Hz$^{-1/2}$ Hz$^{-1/2}$. 

For Fig. 3, we utilize the parameters compatible with the stimulated superradiance pulse experiment at cryogenic temperature \citep{AAngerer2018}: the resonator has the frequency  $\omega_{c}=2\pi\times3.18$ GHz, the photon loss rate $\kappa_{c}=2\pi\times13.8$ MHz  ($\kappa_1=\kappa_{c}/2$), and the coupling with single spin $g_{s}=2\pi\times83.1$ mHz; the spins have the transition frequency  $\omega_{s}=\omega_{c}$, the spin-lattice relaxation rate $\gamma_{s}=2\pi\times0.157$ Hz, and the dephasing rate  $\chi_{s}=2\pi\times {4.7}$ MHz. The number of the spins is  $N=1.5\times10^{16}$, and the driving microwave field drives the resonator resonantly ($\omega_d = \omega_c$) with the field amplitude $\Omega = 2 \pi \times 17\times 10^{10} $Hz$^{-1/2}$ and duration of $28$ ns. 

\section{Hybrid Modes under the Holstein-Primakoff Approximation \label{sec:HPApp}} 

In Fig. \ref{fig:Rabi} of the main text, we show that the collective interaction of the optically cooled spin ensemble with the resonator field leads to Rabi oscillations and cavity mode splittings at room temperature. To gain more insights into the physics, we note that the spin ensemble occupies states near the lower boundary of the Dicke space, which permits us to apply the Holstein-Primakoff approximation \cite{JAGyamfi2019,THolstein,QWu2021} and approximate these states as the occupation number states of quantized harmonic oscillators. As a result, we can  approximate the Hamiltonian of the spin-ensemble as $\hat{H}_{s}\approx\sum_{J} \hat{H}_{s,J}$
with $\hat{H}_{s,J}\approx\hbar\omega_{s}\hat{b}_{J}^{+}\hat{b}_{J}$, where $\hat{b}_{J}^{+},\hat{b}_{J}$
are the bosonic creation and annihilation operator for given $J$  and the spin ensemble-resonator coupling Hamiltonian as $\hat{H}_{a-c}\approx\sum_{J}\hat{H}_{a-c,J}$
with $\hat{H}_{a-c,J}=\hbar\sqrt{2J}g_{s}\left(\hat{a}^{\dagger}\hat{b}_{J}+\hat{b}_{J}^{+}\hat{a}\right)$, where the coupling strength $\sqrt{2J}g_s$ depends on the numbers $J$.Since $\hat{H}_{s}$ is sum of $\hat{H}_{a-c,J}$ over the different $J$, the system response can be veiwed as the sum of response for sub-system with given $J$ weighted by the population on the corresponding Dicke ladder.  For given $J$, we can diagonalize the approximate Hamiltonian $\hat{H}_{s,J}+\hat{H}_{a-c,J}+\hat{H}_{a-c,J}$ as $\sum_{\alpha=\pm}\hbar \omega_\alpha \hat{h}^\dagger_\alpha \hat{h}_\alpha$ with the frequencies $\omega_{\pm}=\frac{1}{2}(\omega_s+\omega_c\pm \chi)$ and the operators
$\hat{h}_\pm =\sqrt{1\pm \frac{\omega_s-\omega_c}{\chi}}[\pm \frac{\hat{b}_J}{\sqrt{2}}+\frac{g_s\sqrt{4J}}{\chi\pm(\omega_s-\omega_c)}\hat{a}]$ of the hybrid modes. Here, the abbreviation is defined as $\chi =\sqrt{8g_s^2J+(\omega_s-\omega_c)^2}$. For the particular case with $\omega_s=\omega_c$, we obtain $
\chi=2\sqrt{2J}g_s$ and $\omega_{\pm}-\omega_c=\pm \sqrt{2J}g_s$.

\begin{figure}[htbp!]
\begin{centering}
\includegraphics[scale=0.50]{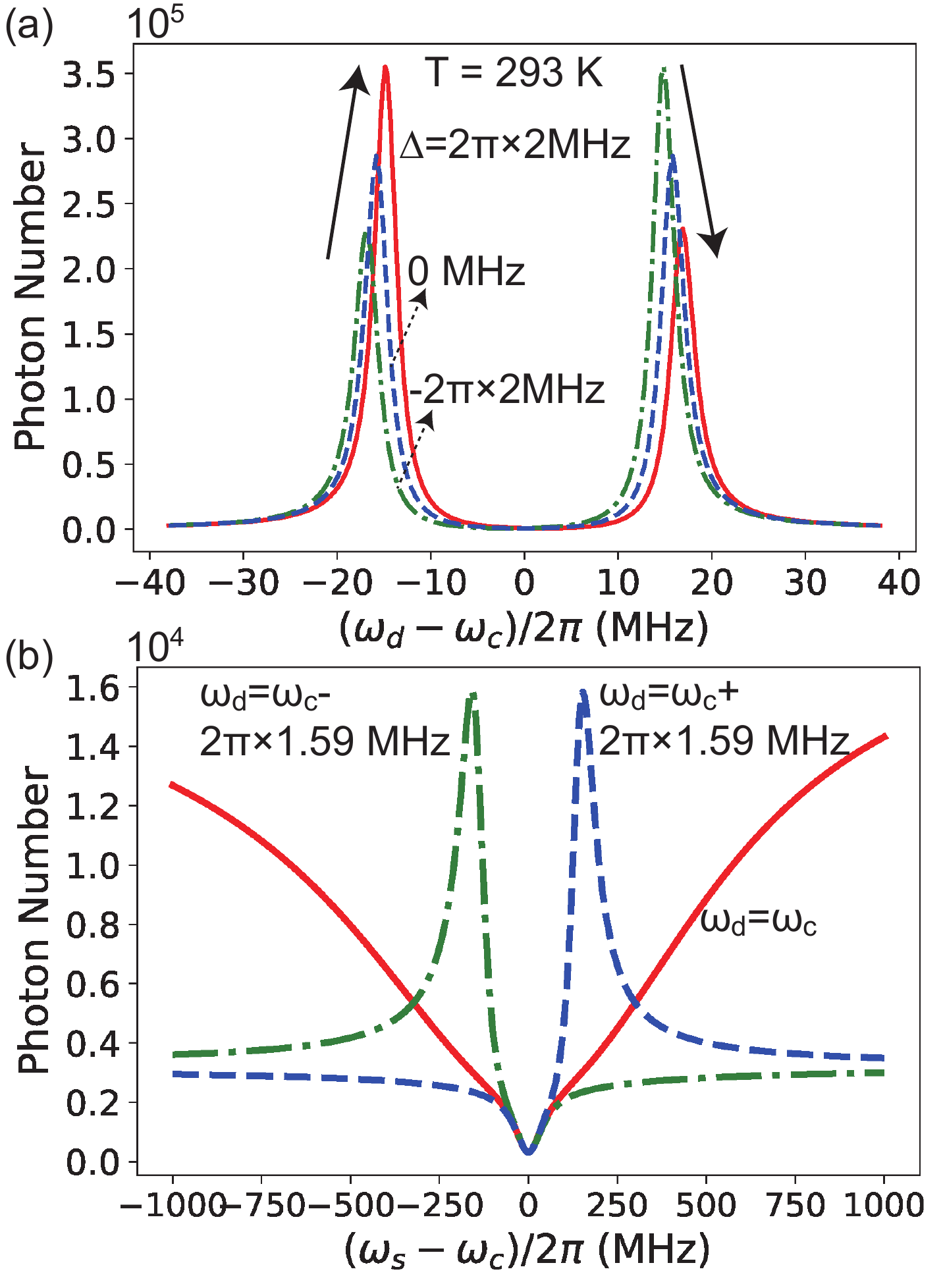}
\par\end{centering}
\caption{\label{fig:cQEDTransition} Sensing with Rabi splitting signals. Panel (a) shows the intra-resonator photon number, and hence the cavity transmission, as function of the frequency detuning $\omega_d-\omega_c$ of the probe field from the resonator for different values of the spin detuning $\Delta=\omega_s - \omega_c$. Here, from the left (green, dot-dashed) curve to the right (blue, dashed and red, solid) ones, the detuning is  $\Delta/(2\pi)=-2,0,2{\rm MHz}$, respectively. Panel (b) shows the photon number, and hence the cavity transmission, as  function of the spin detuning $\Delta=\omega_s - \omega_c$ for $\omega_d = \omega_c$ (red solid line), $\omega_d = \omega_c - 2\pi\times 1.59$ MHz (green dot-dashed line) and $\omega_d = \omega_c + 2\pi\times 1.59$ MHz (blue dashed line). The driving field strength is $\Omega = 2\pi \times 5\times 10^7$ Hz, and the other parameters are the same as those used  in  Fig.\ref{fig:Rabi}(b).}
\end{figure}

\section{Sensing with the Rabi Splitting Signal \label{sec:sensing}} 

In this appendix, we show the dependence of the Rabi splittings on the spin transition frequency $\omega_s$, see Fig. \ref{fig:cQEDTransition}, which can be potentially useful for sensing of magnetic fields and other related quantities. Fig.\ref{fig:cQEDTransition}(a) shows that the peaks blue-shift as $\omega_s$  increases over the resonator frequency, and the maximum of the peaks becomes also stronger. From the analysis in Appendix \ref{sec:HPApp}, we obtain the frequencies of the hybrid modes $\omega_{\pm}=\frac{1}{2}[\omega_s+\omega_c\pm \sqrt{8g_s^2J+(\omega_s-\omega_c)^2}]$ leading to these peaks. For smaller frequency detuning $|\omega_s-\omega_c| \ll 2\sqrt{2J}g_s$, these frequencies can be approximated as $\omega_{\pm}\approx \frac{1}{2}(\omega_s+\omega_c \pm 2\sqrt{2J}g_s)$. The linear scaling of these frequency with $\omega_s$ can be used to infer the change of the spin transition frequency. Furthermore, we note that the sum of the frequencies of the two resonant Rabi peaks $\omega_{+}+\omega_{-} = \omega_s + \omega_c$ are independent of the collective coupling $\sqrt{2J}g_s$ and are also proportional to $\omega_s$. Thus, using the frequency sum, we can infer the spin transition frequency $\omega_s$ more faithfully. 

Besides measuring the frequencies of peaks, we can also determine the change of photon number and hence transmission as a function of the spin transition frequency $\omega_s$ for different values of the frequency $\omega_d$ of the microwave probe field. Such results are shown in Fig. \ref{fig:cQEDTransition}(b), and we see that, when the probe field is resonant with the cavity ($\omega_d=\omega_c$, the red solid line), the photon number is minimal for $\omega_s=\omega_c$ and increases linearly and smoothly with increasing detuning $\omega_s-\omega_c$. In contrast, when the probe field is detuned from the resonator (e.g. $\omega_d-\omega_c = \pm 2\pi\times 1.59$ MHz, the green dot-dashed and blue dashed lines), the photon transmission shows sharp peaks when $\omega_s=\omega_c \pm 2\pi\times 125$ MHz, and relatively broad dips around $\omega_s = \omega_c$. Note that the detuning of these peaks to the resonator is much larger than the dephasing rate induced by the inhomogeneous broadening $\chi=2\pi\times2.6$ MHz. The different behaviors in the resonant and off-resonant case may be beneficial for the wide-band, and narrow-frequency sensing of the spin transition frequency.


\begin{thebibliography}{10}

\bibitem{HMabuchi} H. Mabuchi, A. C. Doherty Cavity quantum electrodynamics: Coherence in context. Science \textbf{298},1372-1377 (2002).

\bibitem{JMRaimond} J. M. Raimond, M. Brune, S. Haroche, Manipulating quantum entanglement with atoms and photons in a cavity. Rev Mod Phys. \textbf{73} (3):565-582 (2001).

\bibitem{JYe} J. Ye, H. J. Kimble, H. Katori, Quantum State Engineering and Precision Metrology Using State-Insensitive Light Traps. Science \textbf{320} (5884):1734-1738 (2008).

\bibitem{RAmsuss2011} R. Ams{\"u}ss, C. Koller, T. N{\"o}bauer, S. Putz,
S. Rotter, K. Sandner, S. Schneider, M. Schramb{\"o}ck, G. Steinhauser, H. Ritsch, J. Schmiedmayer, and J. Majer, Cavity QED with magnetically coupled collective spin states, Phys. Rev. Lett. \textbf{107}, 060502 (2011).

\bibitem{YKubo2010} Y. Kubo, F. R. Ong, P. Bertet, D. Vion, V. Jacques,
D. Zheng, A. Dr{\'e}au, J.-F. Roch, A. Auffeves, F. Jelezko, J. Wrachtrup, M. F. Barthe, P. Bergonzo, and D. Esteve, Strong coupling of a spin ensemble to a superconducting resonator, Phys. Rev. Lett. \textbf{105}, 140502 (2010).

\bibitem{AAngerer2016} A. Angerer, T. Astner, D. Wirtitsch, H. Sumiya, S. Onoda, J. Isoya, S. Putz, and J. Majer, Collective strong coupling with homogeneous Rabi frequencies using a 3D lumped element microwave resonator, Appl. Phys. Lett. \textbf{109}, 033508 (2016).

\bibitem{SPutz2014} S. Putz, D. O. Krimer, R. Ams{\"u}ss, A. Valookaran, T. N{\"o}bauer, J. Schmiedmayer, S. Rotter, and J. Majer, Protecting a spin ensemble against decoherence
in the strong-coupling regime of cavity QED, Nat. Phys. \textbf{10}, 720-724 (2014).

\bibitem{AAngerer2018} A. Angerer, K. Streltsov, T. Astner, S. Putz, H. Sumiya, S. Onoda, J. Isoya, W. J. Munro, K. Nemoto, J. Schmiedmayer,and J. Majer, Superradiant emission from colour centres in diamond, Nat. Phys. \textbf{14}, 1168-1172 (2018). 

\bibitem{JDBreeze2018}J. D. Breeze, E. Salvadori, J. Sathian, N. M. Alford, Kay C. W. M.   Continuous-wave room-temperature diamond maser. Nature. \textbf{555}(7697):493-496 (2018). 
\bibitem{Eisenach2021}E. R. Eisenach, J. F. Barry, M. F. OKeeffe, et al. Cavityenhanced microwave readout of a solid-state spin sensor. Nat Commun. 12(1):1357 (2021).

\bibitem{Ebel2020} J. Ebel, Joas, T. Schalk M., Angere. A. Majer J., Reinhard F., Dispersive readout of room temperature spin qubits, ArXiv:2003.07562v1 (2020).

\bibitem{JDBreeze2017}J. D. Breeze, E. Salvadori, J. Sathian, N. M. Alford, C. W. M. Kay, Room-temperature cavity quantum electrodynamics with strongly coupled Dicke states. npj Quantum Inf. \textbf{3}(1):40 (2017).

\bibitem{RHDicke1054}R. H. Dicke, Coherence in spontaneous radiation processes, Phys. Rev. \textbf{93}, 99 (1954).

\bibitem{YZhang2018} Y. Zhang, Y-X. Zhang, K. M{\o}lmer, Monte-Carlo simulations of superradiant lasing. New J Phys. \textbf{20}(11):112001 (2018). 

\bibitem{BQBaragiola} B. Q. Baragiola, B. A. Chase, J. M. Geremia, Collective uncertainty in partially polarized and partially decohered spin-$\frac{1}{2}$ systems. Phys Rev A. \textbf{81}(3):32104 (2010). 

\bibitem{NShammah2018} N. Shammah, S. Ahmed, N. Lambert, S. De Liberato, F. Nori, Open quantum systems with local and collective incoherent processes: Efficient numerical simulations using permutational invariance. Phys Rev A. \textbf{98} (6):063815 (2018).

\bibitem{JAGyamfi2019}J. A. Gyamfi, An Introduction to the Holstein-Primakoff
transformation, with applications in magnetic resonance, ArXiv:1907.07122 (2019). 

\bibitem{THolstein} T. Holstein and H. Primakoff, Field dependence of the intrinsic domain magnetization of a ferromagnet, Phys. Rev. \textbf{58}, 1098-1113 (1940).

\bibitem{QWu2021} Qilong Wu, Yuan Zhang, Xigui Yang, Shi-Lei Su, Chongxin Shan, Klaus M{\o}lmer, A superradiant maser with nitrogen-vacancy center spins, ArXiv:2105.12350 (2021); Sci. China: Phys. Mech. Astron., Accepted.

\bibitem{KDebnath2018} K. Debnath, Zhang Y, M{\o}lmer K. Lasing in the superradiant crossover regime. Phys. Rev. A. \textbf{98}(6), 063837 (2018). 

\bibitem{YZhang2021} Zhang Y, Shan C, M{\o}lmer K. Ultranarrow superradiant lasing by dark atom-photon dressed states. Phys. Rev. Lett. \textbf{126}(12), 123602 (2021). 

\bibitem{HAMLeymann2014} H. A. M. Leymann, A. Foerster, and J. Wiersig, Expectation value based equation-of-motion approach for open quantum systems: A general formalism, Phys. Rev. B \textbf{89}, 085308 (2014).

\bibitem{DPlankensteiner2021}D. Plankensteiner, C. Hotter, H. Ritsch, QuantumCumulants.jl: A Julia framework for generalized mean-field equations in open quantum systems, ArXiv:2105.01657 (2021)


\bibitem{KDebnath2020} K. Debnath, G. Dold, J. J. L. Morton, K. M{\o}lmer, Self-stimulated
pulse echo trains from inhomogeneously broadened spin ensembles. Phys. Rev. Lett.  \textbf{125}(13):137702 (2020). 

\bibitem{IDiniz2011} I. Diniz, S. Portolan, R. Ferreira, J. M. G{\'e}rard,
P. Bertet, A. Auff{\`e}ves, Strongly coupling a cavity to inhomogeneous ensembles
of emitters: Potential for long-lived solid-state quantum memories. Phys. Rev. A \textbf{84}(6):063810 (2011). 

\bibitem{JWesenberg} J. Wesenberg, K. M{\o}lmer, Mixed collective states of many spins. Phys Rev A. \textbf{65}(6):62304 (2002). 

\bibitem{CJWood} C. J. Wood, D. G. Cory, Cavity cooling to the ground state of an ensemble quantum system. Phys Rev A. \textbf{93} (2):23414 (2016).  

\bibitem{ABychek} A. Bychek, C. Hotter, D. Plankensteiner, H. Ritsch, Superradiant lasing in inhomogeneously broadened ensembles with spatially varying coupling, arXiv:2105.11023 (2021)

\end{thebibliography}
\end{document}